\documentstyle[epsfig, twocolumn]{esapub}

\begin{document}

\setlength{\parindent}{0pt}
\setlength{\parskip}{ 10pt plus 1pt minus 1pt}
\setlength{\hoffset}{-1.5truecm}
\setlength{\textwidth}{ 17.1truecm }
\setlength{\columnsep}{1truecm }
\setlength{\columnseprule}{0pt}
\setlength{\headheight}{12pt}
\setlength{\headsep}{20pt}
\pagestyle{esapubheadings}

\title{\bf MULTIWAVELENGTH OBSERVATIONS OF GAMMA-RAY BURSTS: \\
 TOWARDS THE UNDERSTANDING OF THE MYSTERY}

\author{{\bf Alberto J.~Castro-Tirado} \vspace{2mm} \\
Laboratorio de Astrof\'{\i}sica Espacial\\
 y F\'{\i}sica Fundamental (LAEFF-INTA)\\
P.O. Box 50727, E-28080, Madrid, Spain \\
e-mail address: ajct@laeff.esa.es \\}

\maketitle

\begin{abstract}

Since their discovery in 1973, Gamma-Ray Bursts (GRBs) have remained for many
years one of the most elusive mysteries in High Energy-Astrophysics.
The main problem regarding the nature of GRBs has usually been the lack of
knowledge of their distance scale. About 300 GRBs are detected annua\-lly by 
BATSE in the full sky, but only a few of them can be loca\-lized accurately
to less than half a degree. For many years, follow-up observations
by other satellites and ground-based telescopes were conducted, 
but no counterparts at other wavelengths were found. The breakthrough took 
place in 1997, thanks to the observation by {\it BeppoSAX} and {\it RossiXTE} 
of the fa\-ding X-ray emission that follows the more energetic 
gamma-ray photons once the GRB event has ended. 
This emission (the afterglow) extends at longer wavelengths, and 
the good accuracy in the position determination by {\it BeppoSAX} has led to 
the discovery of the first optical counterparts -for GRB 970228, GRB 970508, 
and GRB 971214-, greatly improving our understanding of these 
puzzling sources.
Now it is widely accepted that most bursts originate at cosmological
distances but the final solution of the GRB problem is still far away.
\vspace {5pt} \\


  Key~words: multiwavelength observations; gamma-ray bursts.

\end{abstract}

\section{INTRODUCTION}

In 1967-73, the four {\it VELA} spacecraft (named after the spanish verb 
{\it velar}, to keep watch), that where originally designed for verifying 
whether the former Soviet Union abided by the Limited Nuclear Test Ban 
Treaty of 1963, observed 16 peculiarly strong events 
(Klebesadel, Olson and Strong 1973, Bonnell and Klebesadel 1996).
On the basis of 
arrival time differences, it was determined that they were related neither
to the Earth nor to the Sun, but they were of cosmic origin. Therefore they
were named cosmic $\gamma$-ray Bursts (GRBs hereafter).

GRBs appear as brief flashes of cosmic high energy photons, emitting the 
bulk of their energy above $\approx$ 0.1 MeV (Fig. 1). 
They are detected
by instruments somewhat similar to those used by the particle physicists at
their laboratories. The difference is that GRB detectors have to be placed
onboard balloons, rockets or satellites.
In spite of the abundance
of new observations of GRBs, their energy source and emission mechanism remain
highly speculative. 

The KONUS experiment on {\it Veneras 11} and {\it 12} gave the first 
indication that GRB sources were isotro\-pically distributed in the sky 
(Mazets et al. 1981, Atteia et al. 1987).  Based on a much larger sample, this 
result was nicely confirmed by BATSE on board the {\it CGRO} (Meegan et al. 
1992). About 300 GRBs occur annually in the full sky, but only
few of them are localized accurately.  The apparent isotropy 
was interpreted in terms of GRBs produced at cosmological distances, 
although the possibility of a small fraction of the sources lying nearby, 
within a galactic disc scale of few hundred pc, or in the halo of the Galaxy, 
could not be discarded. Another result was that the time profiles of the bursts
are very diffe\-rent, with some GRBs lasting a few ms and others lasting for 
several minutes.  In general, there was no evidence of periodicity in the 
time histories of GRBs.  However there was indication of a bimodal 
distribu\-tion of burst durations, with
$\sim$25\% of bursts ha\-ving durations around 0.2 s and $\sim$75\% with 
durations around 30 s. A extensive review of the observational 
character\-istics can be found in Fishman and Meegan (1995).

A deficiency of weak events was also noticed, and all these observational
data led many researchers to believe that GRBs are indeed at cosmological
distances.  In this case, the released energies could
be as high as 10$^{53}$ erg and models could involve coalescence of neutron
stars in double systems, neutron star-black hole systems, accretion
induced collapse in white dwarfs or 
$^{\prime\prime}$failed$^{\prime\prime}$ Type I supernovae.
It was also proposed the possibility of having a mixture of two popu\-lations: 
a cosmological plus a galactic one, but the latter, probably formed by
accreting neutron stars, will account for only a very small fraction of the 
total population. See Nemiroff (1984) for a review of the di\-fferent 
theoretical models.

It is well known that an
important clue for resol\-ving the GRB puzzle is the detection of transient
emission -at longer wavelengths- associated with the bursts. 
Here I review all the efforts in the search for GRB counterparts throughout  
the electromagnetic spectrum. 
I will first review the searches prior to 1997, and afterwards I will 
discuss the important discove\-ries achieved last year. 
Previous reviews can be seen in Schaefer (1994), Hartmann (1995), Vrba (1996),
Greiner (1996a) and Hurley (1998).

\begin{figure}[htp]
\epsfxsize=70mm
\epsfysize=60mm
\centerline{\epsfbox{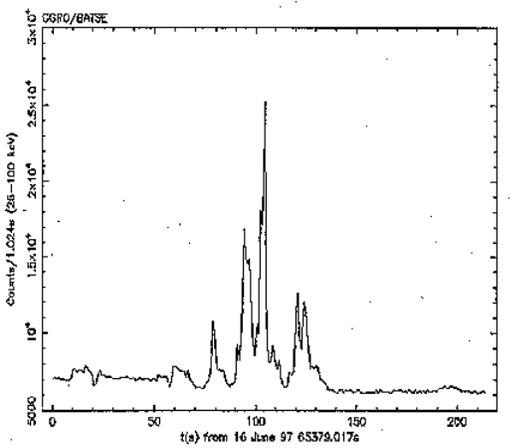}}
\caption{One of the GRBs detected by {\it BATSE}, lasting for about
150-s. Further details in Connaughton et al. (1997).}
\end{figure}

\section{SEARCHES PRIOR TO 1997}

\subsection{X-ray searches}

\subsubsection{Searches for simultaneous and near-simultaneous X-ray emission}

The observation of X-rays below 20 keV is important in order to understand 
the nature of the phy\-sical processes in the GRB sources. Most of the 
experiments devoted to GRB study had lower cut-off energies of the order of
20-40 keV. However, a few of them were designed in order to cover part of 
the soft X-ray range.

Already in 1972, the Apollo 16 spectrometer detected X-ray photons in the
2-8 keV range for GRB 720427 (Trombka et al. 1974). 
Wheaton et al. (1973) observed in GRB 720514 the presence of an
X-ray signal which lasted {\it longer} than the gamma-ray event.
In 1984, Katoh et al. reported an X-ray tail for GRB 811016 that lasted more
than 30 s after the termination of the hard event. 
{\it GINGA} also detected emi\-ssion down to 1.5 keV, with soft X-ray tails 
found for 8 events (Murakami et al. 1991, Yoshida and Murakami 1994).
Moreover, a significant fraction ($\sim$ 10\%) of the GRBs detected by WATCH 
on {\it GRANAT} displayed X-ray tails in the 6-20 keV range 
(Castro-Tirado et al. 1994a).  Connors and McConnell (1996) have discussed 
the proposal that the X-ray transient detected by {\it HEAO 1} following GRB 
780506 could indeed be related to the burst itself.
Therefore it seemed clear that, at least for some bursts, there was evidence for X-ray emission lasting longer than the GRB.

\subsubsection{Deep searches}

Grindlay et al. (1982) and Pizzichini et al. {1986} initiated the searches 
for X-ray quiescent 
counterparts by means of {\it EINSTEIN}. Five GRB error boxes were observed 
and, with one possible exception (GRB 781119, Grindlay et al. 1982), no point 
source was detected in any of the observations. Assuming
that the bursts originated from galactic neutron stars, only upper limits
to the surface temperatures were derived (Pizzichini et al. 1986).
Boer et al. (1988) observed the GRB 781119 region, but failed to detect the 
source 
with {\it EXOSAT}. Recently, {\it ASCA} and {\it ROSAT} "rediscovered" the 
X-ray source in GRB 781119, with  a flux of 
F$_{(0.5-8 \ \rm keV)} \sim$ 1.6 $\times$ 10$^{-13}$ erg cm$^{-2}$ s$^{-1}$ 
 (Hurley et al. 1996, Boer et al. 1997). 
Further evi\-dence for an X-ray source associated with a GRB was obtained 
by {\it ASCA} in the GRB 920501 error box (Murakami et al. 1996). The X-ray
spectrum is heavily absorbed at low energies, implying an absorption consistent
with the galactic value in that direction, and sugges\-ting that the X-ray
source is {\it quite} distant. Using the {\it ROSAT} all-sky-survey data, 
Greiner et al. (1995) detected 27 X-ray
sources in 45 small GRB error boxes. On the basis of the optical identification of theses sources, no convincing association to GRBs was found. Moreover, the 
number of sources corresponded to the expected number of background sources. 
The upper limit to the X-ray flux was 
F$_{(0.1-2.4 \ \rm keV)} \sim$  10$^{-13}$ erg cm$^{-2}$ s$^{-1}$. 
Ano\-ther field (GRB 940301) was observed by {\it ROSAT} four weeks later and
revealed no fading counterpart (Greiner et al. 1996b).

\subsection{EUV searches}

\subsubsection{Near simultaneous searches}

Searches at EUV wavelengths have only been performed since 1992, following
the launching of {\it EUVE}.
Castro-Tirado et al. (1998a) has reported a serendi\-pitous observation in the
extreme ultraviolet window
by {\it EUVE}, $\sim$ 11 hours after the occurrence of GRB 921013b. 
No source was detected in the EUV range.

\subsubsection{Deep searches}

A deep observation on the GRB 920325 error box was performed 17 months after
by {\it EUVE}. No quiescent source was found, thus constraining a neutron
star as a possible counterpart (Hurley et al. 1995).

\subsection{Optical searches}

\subsubsection{Archival searches}

The strategy behind this kind of search is the assump\-tion that GRBs
repeat.  Some scientists have assumed that GRBs could be a repetitive  
pheno\-menon, and they have used all the available information at the major 
plate archives in order to look for
optical transient (OT) emission in some of the smaller GRB error boxes. 
Atteia et al. (1985) searched for about 1500 hours but no such transients were
seen. Schaefer discovered in 1981 the first OT in a Harvard plate taken in
1928 (OT 1928), located inside the small GRB 781119 error box. Another OT 
was found in the GRB 781006b error box in three different plates taken in
1966. A large amplitude flaring dMe star, probably unrelated to the burst,
was proposed to be related to the OT (Greiner and Motch 1995).
Ano\-ther good candidate is OT 1905, found by Hudec et al. (1994) slightly
outside the interplanetary network error box for GRB 910219. 
Although these OTs seem to be real, another $\sim$ 40 candidates were 
found, but most of them were rejected as they turned out to be plate 
defects. See a discussion in Hudec (1993, 1995). However,
there is no convincing proof that the OTs found in plate archives and GRBs 
are related to the same physical phenomenon.

\subsubsection{Deep searches}

Deep searches were carried out in the 80's for the 
smallest GRB error boxes provided by the Interplanetary Network: GRB 781119 
(studied by Pedersen et al. 1983) and GRB 790406 (Chevalier et al. 1981). 
Recently, with the advent of the newer big CCD detectors, deep searches have 
been carried out of GRB 790329 and other error boxes (Vrba, Hartmann and 
Jennings 1995). The most powerful telescopes, such as the 6-m in the 
Caucasus and the Hubble Space Telescope (HST) have
also been used for observing the error boxes of GRBs: GRB 790613 
(Sokolov et al. 1996)
and GRB 790325 and 920406 (Schaefer et al. 1997a). 
In any case, no quiescent GRB counterpart has been firmly established.
In general, the small GRB error regions did not contain galaxies to fairly
deep limits (the so-called $^{\prime\prime}$no-host problem$^{\prime\prime}$, 
Fenimore et al. 1993), and it was obvious that 
the bursts, if at cosmological distances, could not come from normal galaxies. 
Either they had to occur in subluminous galaxies or their progenitors somehow 
had to be ejected from the parent galaxy.

\subsubsection{Near-simultaneous searches}

The first search for optical emission arising {\it simultaneously} with a
GRB was performed by Grindlay, Wright and McCrowsky (1974). Hudec et al. (1987) and Hudec (1993) 
found five events for which time-correlated plates were available. 
Greiner et al. (1996c) found plates within $\pm$ 12-h of the bursts, but none 
revealed 
interesting brightenings in the error box. For GRB 920824, a rather deep plate 
was taken simultaneously with the burst in Dushanbe. 
They conclude that the optical emission of typical GRBs is at level below
 $F_{opt}/F_{\gamma}$ $\sim$ 2 {\it during} the burst and 
 $F_{opt}/F_{\gamma}$ $\sim$ 20-400 a
few hours {\it after} the burst, with $F_{\gamma}$ and $F_{opt}$ being the peak 
gamma-ray (and optical) fluxes.
Stronger upper limits have been obtained by GROSCE (Lee et al. 1996) who
have recorded sky images for more than 30 GRB triggers, in many cases while 
the burst was still in progress.
Additional results were obtained by Castro-Tirado et al. (1994b) regarding
follow-up observations of eight bursts detected by WATCH. No transient optical
emission was detected in the fo\-llowing hours down to 18th magnitude.
Similar observations were reported for GRB 930131 by Schaefer et al. (1994)
and McNamara and Harrison (1994). Again, no firm counterpart was detected in
their observations.  Barthelmy et al. (1994) obtained upper limits for any 
transient optical emission arising from three bursts, between 35 hr and 8 days 
following the burst.
The only OT found nearly-simultaneously to a GRB event, is the one reported by
Borovi\u{c}ka et al. (1992). They found a star-like spot at the edge of the
GRB 790929 error box on a plate taken only 7.1 hr after the high energy
event.  Also, two additional objects were found at the same position on
another plates.  The objects are consistent with the star HDE 249119.
 Further optical studies detected additional optical eruptions of HDE 249119,
reporting ten optical brightenings with amplitudes greater than 0.5 mag
(\u{S}t\u{e}p\'an \& Hudec 1996), but it is unlikely that this star would be 
related to GRB 790929.

\subsection{Near-IR searches}

\subsubsection{Deep searches}

The first search for counterparts in the near-IR was initiated by 
Schaefer et al. (1987), who investigated seven GRB error boxes at a 
wavelength of
2.2 $\mu$m (K-band). No convincing candidates were found, with a detection 
limit of K = 19 for two error boxes. An imaging survey of the six smallest 
GRB error boxes from the third Interplanetary
Network has revealed relatively bright galaxies in each of them (Larson,
McLean and Becklin, 1996). Although the statistical significance of their 
result is not overwhelming, it leaves the "no-host problem" as an open 
question.
Klose et al. (1996) obtained deep imaging of the GRB 790418 error box, showing 
several galaxies as well.
Deep infrared imaging of the quiescent X-ray source located within the GRB
920501 error box has been performed by Blaes et al. (1997). One of the
objects is a type 1 Seyfert galaxy, and Drinkwater et al. (1997) has 
identified the galaxy 
with the X-ray source seen by {\it EINSTEIN}, {\it ASCA} and {\it ROSAT}, 
strengthening the above-mentioned results. Is this Seyfert galaxy
the true GRB counterpart?

\subsection{Mid-IR searches}

Searches for {\it quiescent} counterparts have also been conducted in the 
past in the IR. Schaefer et al. (1987) used the IRAS
data base at wavelenghts of 12, 25, 60 and 100 $\mu$m and looked for candidates
within 23 well localized GRB error boxes. No convincing
counterparts were found.

\subsection{Millimetre searches}

Two simultaneous observations were obtained by the COBE satellite,
when correlating GRBs that o\-ccurred during an 8 month overlap period 
(Bontekoe et al. 1995, Stacy et al. 1996).
The 2$\sigma$ upper limit quoted is 31000 Jy at 53 GHz.
A likelihood analysis was performed by Ali et al. (1997) in order to look 
for a change in the level
of millimetric emission from the location of 81 GRBs during the first
two years of the mission (1990-91). No positive detection was achieved,
with 95 \% confidence level upper limits of 175, 192 and 645 Jy respectively 
at 31, 53 and 90 GHz.
Another team at the Fallbrook Low-Frequency Immediate Response
Telescope (FLIRT) searched for prompt radio emission at 74 MHz arising 
from GRB sources. The delays were of the order of 2 to 12 min. Preli\-minary
results for 12 GRBs indicated that no source was found above a $\sim$ 100
Jy detection limit (Balsano et al. 1996).

\subsection{Radio searches}

\subsubsection{Near simultaneous searches}

A serendipitous observation towards the error box of GRB 920711 was
performed at 151 MHz with the Cambridge Low-Frequency Synthesis Telescope
(CLFST) from 2 to 4 days after the occurrence of GRB 920711.
No source was detected down to 40 mJy (Koranyi et al. 1994). 
This instrument also observed GRB 940301 within 1 hour after the burst, 
imposing a 200 Jy upper limit, and 80 mJy between 1-36 days after the event. 
CLFST observations were conducted for another two bursts, with
flux limits of 35 Jy for 2.5 hours after GRB 950430 and 16 Jy for 1 hour 
after GRB 950706 (Green et al. 1996). 
A search for a radio transient at 4.86 and 1.4 GHz within the GRB 930131
failed to detect any source (Schaefer et al. 1994).
VLA observations were performed 23 hours after GRB 940217, but only an
upper limit of  $\leq$ 4 $\mu$Jy/beam was given by Palmer et al. (1985).
No fading/flaring radio counterpart was found either at 1.4 GHz (upper limit of
3.5 mJy) or 0.4 GHz (upper limit of 55 mJy) within the GRB 940301 error
box (Frail et al. 1994).

\subsubsection{Deep searches}

It was suggested that if GRBs were to be related to highly magnetized, rotating
neutron stars, they would produce polarized radio emission during the 
quiescent state.
In fact, VLA maps of the GRB 781119 error box showed two sources (Hjellming 
and Ewald 1981). One showed some degree of polarization, and the other 
coincided with the {\it ROSAT} X-ray source (Hurley et al. 1996). 
Schaefer et al. (1989) found four radio sources in 10 small GRB error regions. 
None of them displayed unusual properties and the number of sources was the 
expected number of chance background sources. They concluded that none of
them were associated with GRBs.
Deep VLA observations of three GRBs were also performed by Palmer et al. (1995). 
GRB 920501 was observed 13 days and 9 months after the burst, at frequencies 
of 1.4 and 8.4 GHz, imposing upper limits of $\leq$ 
235 $\mu$Jy/beam and $\leq$ 77 $\mu$Jy/beam respectively. A strong
limit was derived for GRB 930706 as $\leq$ 1 $\mu$Jy/beam after 9 days.

\subsection{VHE/UHE searches}

Photons with energies up to 18 GeV have been detected in only one case, 
1.5-hr after GRB 940217 (Hurley et al. 1994). There are models that predict 
TeV emission from GRBs (Meszaros, Rees and Papathanassiou 1994, Halzen et al. 
1991).  A po\-sitive detection would constrain models with respect to the 
emission mechanism and the distance, because no TeV photons are expected if 
they occur at cosmological distances due to the interaction of the photons 
with the microwave and infrared background. This 
process leads to pair production and might become important at E $\geq$ 30 
TeV, as pointed out by Wdowczyk, Tkaczyk and Wolfendale (1992), Mannheim, 
Jartmann and Funk (1997) and Stecker and de Jager (1997).
In fact, at E $\geq$ 50 TeV, there seems to be evidence of a detection from 
the Dublin ASA (GRB 910511, Plunkett et al. 1995). 
The rest of the current arrays and Cerenkov telescopes have provided only upper 
limits at this e\-nergy threshold. 
A systematic search was conducted at AES-KGF (Bhat et al. 1996) for PeV 
$\gamma$-rays from the locations of about 39 high fluence GRBs du\-ring 
Apr 91-Mar 93. Upper limits of 2 $\times$ 10$^{-12}$ photons 
cm$^{-2}$ s$^{-1}$ were obtained.
Another search was performed for 115 GRBs at the CASAMIA detector. Flux limits
of 6 $\times$ 10$^{-12}$ photons cm$^{-2}$ s$^{-1}$ at E $\geq$ 160 TeV
were obtained (Catanese et al. 1996). 
Follow-up searches have also been carried out with HEGRA above
1 TeV for more than 150 bursts, but no significant excess was detected 
by Krawczynski et al. (1996) or Padilla et al. (1997). 
In the 0.25-4 TeV range, it has been su\-ggested that the Whipple Observatory 
can detect, under favourable conditions, $\sim$ 1 event per year (only 
$\sim$ 0.01 per year in the case of the HEGRA array). 
However, only upper limits have been set up by the Whipple telescope 
(Connors and McConnell 1996)
and the EAS-TOP ASA experiment at E $\geq$ 10 TeV (Agglietta et al. 1995).

\subsection{Neutrinos and muons}

The search for neutrinos, antineutrinos and muons arising either from 
intrinsic GRB mechanisms as proposed by M\'eszaros and Rees (1993) and 
Paczynski and Xu (1994) 
or Earthrock neutrino-produced 
muons, has been performed at several places, but none of the experiments 
have detected neutrinos or muons in excess of the expected background in 
correlation with a GRB. These are the Mount Blanc LSD telescope 
(Agglietta et al. 1995), the Irvine-Michigan-Brookhaven (IMB) detector 
(LoSecco 1994, Becker-Szendy 1995) and the Soudan-2 muon experiment 
(De Muth et al. 1994).  Upper limits are 6 $\times$ 
10$^{11}$ ~cm$^{-2}$ (for ${\nu}_{e}$, ${\nu}_{\mu}$, 
${\nu}_{\tau}$ at 20 $\leq$ E$_{\nu}$ $\leq$ 100 MeV) and  2.1 $\times$
10$^{13}$ ~cm$^{-2}$ (for $\bar{\nu}_{e}$, $\bar{\nu}_{\mu}$, 
$\bar{\nu}_{\tau}$ at 20 $\leq$ E$_{\nu}$ $\leq$ 100 MeV) 
(Agglietta et al. 1995).

\section{1997: THE DETECTION OF THE FIRST COUNTERPARTS AT OTHER WAVELENGTHS}

\subsection{The X-ray afterglow}

With the launching of {\it BeppoSAX}, it was possible on 28 Feb 
1997 to detect the first {\it clear} evidence of a long X-ray tail 
 -the X-ray afterglow- fo\-llowing GRB 970228. See figure 2. 
The X-ray fluence was $\sim$ 40 \% of the gamma-ray fluence, as reported 
by Costa et al. (1997). Therefore, the X-ray 
afterglow is not only the low-energy tail of the GRB, but also a significant 
channel of energy dissipation of the event on a completely different timescale.
Another important result was the non-thermal origin of the burst radiation 
and of the X-ray afterglow (Frontera et al. 1998).

\begin{figure}[tp]
\epsfxsize=80mm
\epsfysize=67mm
\centerline{\epsfbox{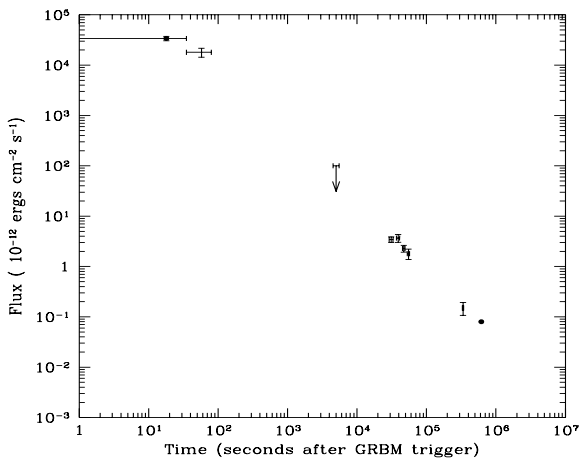}}
\caption{The X-ray afterglow of GRB 970228. Based on the WFC, BeppoSAX and 
ASCA data. Adapted from Costa et al. (1997).}
\end{figure}

Further X-ray afterglows were observed by {\it BeppoSAX} in GRB 970402 
(Piro et al. 1997a), GRB 970508 (Piro et al. 1997b), GRB 971214 
(Antonelli et al. 1997) and probably in GRB 970111 (Feroci et al. 1998) 
and GRB 971227 (Sofitta et al. 1997). Another three were observed
by {\it RossiXTE} in GRB 970616 (Marshall et al. 1997a), GRB 970815 
(Smith et al. 1997) and GRB 970828 (Marshall et al. 1997, Remi\-llard et al. 
1997). The latter was also detected
by {\it ASCA} (Murakami et al. 1997) and {\it ROSAT} (Greiner et al. 1997). 
The X-ray spectrum as observed by {\it ASCA} is strongly absorbed  
(N$_{\rm H}$ = 10$^{22}$ cm$^{-2}$, 
Murakami et al. 1998), suggesting that the event o\-ccurred in a dense medium.

\subsection{A search for EUV counterparts}
Bo\"er et al. (1997b) reported an observation by EUVE $\sim$ 20 hr after 
GRB 971214.  No source was detected at 100 $\rm \AA$. 
Assuming N$_{\rm H}$ = 10$^{20}$ cm$^{-2}$, 
the upper limit to the flux is  1.7 $\times$ 10$^{-13}$ erg cm$^{-2}$ s$^{-1}$.

\begin{figure}[htp]
\epsfxsize=65mm
\epsfysize=65mm
\centerline{\epsfbox{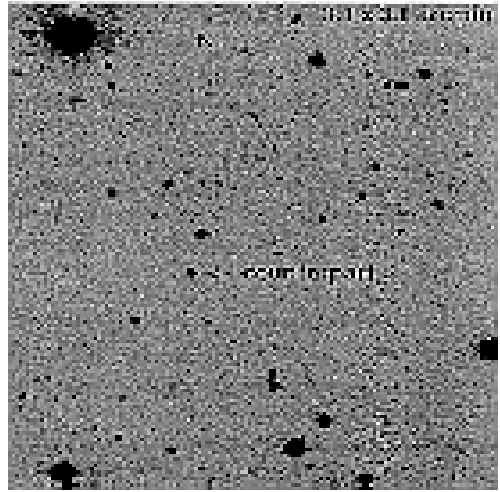}}
\caption{The OT related to GRB 970508, as seen in this R-band image taken
by C. Wolf and R. Fockenbrock with the 2.2m telescope at the German-Spanish
Calar Alto Observatory on 9 May 1997. North is up and east to the left.}
\end{figure}

\subsection{The optical counterparts}

The improvement of GRB localizations by {\it BeppoSAX} (in combination with 
the {\it Third Interplanetary Network} of satellites) has led to the 
detection of the first three optical counterparts in 1997.
The first optical transient asso\-ciated to a GRB was identified on 28 Feb
1997, 20 hr after the event, by Groot et al. (1997).  
The OT was seen in earlier images taken by Pedersen et al. (1997) and 
Guarnieri et al. (1997), 
in the rising phase of the light curve. The maximum was reached $\sim$ 20 hr 
after the event (V $\sim$ 21.3), and followed by a power-law decay as 
t$^{-1.2}$ (Galama et al. 1997, Masetti et al. 1998). 
An extended source was seen at the OT position since the very beginning 
by both ground-based and {\it HST} observations (van Paradijs et 1997, 
Sahu et al. 1997).
New {\it HST} observations taken 6 months after the event were reported by 
Fruchter et al. (1997). Both the OT (at V = 28) and the extended source 
(V = 25.6) were seen. 
The extended source surroun\-ding the point-source was  interpreted as a 
galaxy, a\-ccording to the similarities (apparent size, magnitude) with 
objects in the {\it HST} Deep Field.

The second optical transient is associated with GRB 970508 and was discovered 
by Bond (1997). It was observed only 3 hr after the burst in unfiltered 
ima\-ges (Pedersen et al. 1998a) 
and  4-hr after in both the U and R-bands (Fig. 3). It displayed a strong 
UV excess (Castro-Tirado et al. 1998, Galama et al. 1998). The optical light 
curve reached a peak two days after (R = 19.7, Djorgovski et al. 1997) and 
was followed by a power-law decay as t$^{-1.2}$ (Fig. 4). The long-lasting 
OT was observed
at many ground-based observatories (Chevalier et al. 1997, Garcia et al. 1997,
Mignoli et al. 1997, Schaefer et al. 1997b, Sokolov et al. 1998). Optical 
spectroscopy was performed at Calar Alto and La Palma (Castro-Tirado et al. 
1997a) and Hawaii (Metzger et al. 1997a). The la\-tter spectrum allowed a 
direct determination of the GRB 970508 redshift
($z \geq 0.835$) and was the first proof that at least a fraction of the GRB
sources lie at cosmological distances (Metzger et al. 1997b). Unlike GRB 
970228, no host galaxy has been seen so far, but the optical spectroscopy and 
the 
flattening of the decay in late August (Pedersen et al. 1998b) suggest that 
the contribution of a constant brightness source -the host galaxy- has been
detected. The luminosity of the galaxy is well below the knee of the 
galaxy luminosity function, $L^{*}$, and the detection of deep Mg I 
absorption and strong [O II] emission -both indicative of a dense medium-
suggests that the host could be a dwarf, blue, rapidly 
forming starburst galaxy (Pian et al. 1998).

\begin{figure}[htp]
\epsfxsize=90mm
\epsfysize=67mm
\centerline{\epsfbox{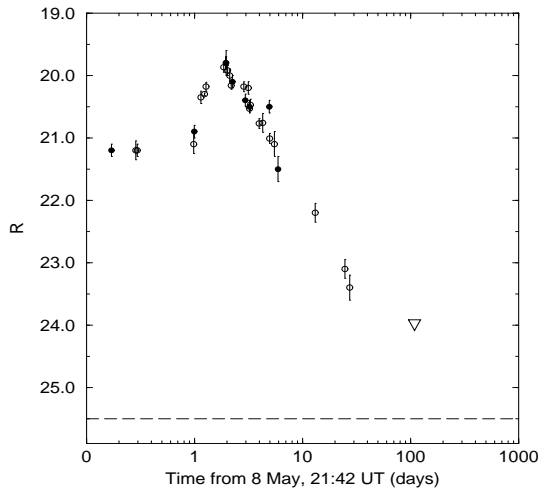}}
\caption{The light curve of OT/GRB 970508 in the R-band. Based on observations
obtained at La Palma, Calar Alto and Loiano. The horizontal dashed line 
indicates the magnitude of the constant source (the galaxy) that is believed to
host the GRB. Adapted from Castro-Tirado et al.  (1998a).}
\end{figure}

The third optical transient is related to GRB 971214 and was identified
by Halpern et al. (1997) as an I = 21.2 object that faded 1.4 mag in 1
day. Independently, this object was also noticed as suspicious by Itoh et 
al. (1997). Further observations proved that the decay followed a power-law 
decline (Diercks et al. 1997, Castander et al. 1997) with t$^{-1.4}$, 
similar to, but steeper than, the other two GRBs with optical counterparts. 

In spite of intensive searches, no optical counterparts were found within
one day of the event for GRB 970111, GRB 970402 or GRB 970828 (Castro-Tirado
et al. 1997b, Pedersen et al. 1998b, Groot et al. 1998).

\subsubsection{Near-infrared counterparts}

Fading near--IR counterparts were observed for GRB 970228, GRB 970508 and 
GRB 971214. For GRB 970228, the IR-flux declined from J = 21.0 
(K $\geq$ 19.5) on Mar 18 (Klose and Tuffs 1997) to J = 23.5 (K = 22.0) on 
Mar 30 (Soifer et al. 1997). 
In the case of GRB 970508, the IR flux decayed from K = 18.4 
to K = 19.2 in three days (Morris et al. 1997).
GRB 971214 was detected in the K-band 4 hr after the burst (Gorosabel et al. 
1998) and decreasing from J = 20.3 on Dec 15 to J = 21.5 on Dec 16 (Tanvir 
et al. 1997).

\subsection{A search for mid-IR counterparts}

Following the detection by BeppoSAX of GRB 970402, a Target-of-Opportunity
Observation by {\it ISO} was performed 55-hours after the event,
in the ISOCAM 8-15 $\mu$m (IRAS-equivalent bandpass) and ISOPHOT 174 $\mu$m 
band, in order to monitor the entire BeppoSAX GRB error box. A further 
observation was carried out 8 days later, with a similar set-up, in 
the X-ray afteglow error box. About 50 sources
were detected in a field centered on the GRB error box. Seven of
them lie within the smaller X-ray error circle. With the exception of the
variable star BL Cir, no other variable object was found in the entire GRB 
error box. 5-$\sigma$ upper limits for any new object are
F$_{(12 \rm\mu m)}$ $\leq$ 0.14 mJy and F$_{(174 \rm \mu m)}$ $\leq$ 350 mJy 
(Castro-Tirado et al. 1998b). A negative result was also obtained in the 
same bandpass for GRB 970508 (Hanlon et al. 1998).

\subsection{Detection in the millimetre range}

GRB 970508 was detected as a continuum point source at 86 GHz with the
IRAM Plateau de Bure Interferometer. A $\sim$ 2 mJy flux was observed 
on 19-21 May 1997. The object was 
not detected any more after 28 May. No significant intra-day variation was 
seen, and it was excluded that the flux was modulated significantly by 
interstellar scintillation (Bremer et al. 1998).
In the case of GRB 971214, no source was detected at 850 $\mu$m at the
James Clerk Maxwell Telescope. The upper limit was 1 mJy on Dec 17-22 
(Smith and Tilanus 1998).

\subsection{Detection of a radio counterpart}

After two unsuccesful searches for transient radio emission in GRB 970111 
and GRB 970228 (Frail et al. 1997a), prompt VLA observations of the GRB 
970508 error box allowed detection of a variable radio source within the error 
box, at 1.4, 4.8 and 8.4 GHz. It coincided with 
the optical transient and displayed a bizarre behaviour, in the sense
that it remained at a steady level of brightness, flaring occasionally by a
factor of two or three (Frail et al. 1997b), suppor\-ting 
the idea of the host galaxy as an active galactic nucleus. 
VLBI observations did not resolve the source (Taylor et al. 1997).
Frail et al. proposed that the fluctuations could be the result of 
strong scattering by the irregularities in the io\-nized Galactic 
interste\-llar gas. 
The damping of the fluctuations with time indicates that the source 
expanded to a significantly larger size. The source was also detected at
15 GHz (Pooley and Green 1997).

All these observational efforts have provided the complete picture of a
GRB multiwavelength spectrum, in the case of GRB 970508 (Fig. 5).

The observational characteristics of the three GRB counterparts (or 
{\it hypernovae}, a name proposed by Paczynski) can be
accommodated in the framework of the fireball models (M\'esz\'aros et al. 1994, 
M\'esz\'aros and Rees 1997, Tavani 1997, Vietri 1997, Waxman 1997, 
Wijers et al. 1997), in which a compact source releases
10$^{53}$ ergs of energy within dozens of seconds. The plasma accelerates
to relativistic velocities (the fireball). The blast wave is moving ahead
of the fireball, and sweeps up the interstellar matter, producing an afterglow
at frequencies gradually declining from X-rays to visible and radio wavelenghts.
Although some of the predicted behaviour has been observed, there are still
many open questions (see Katz and Piran 1998).

\begin{figure}[tp]
\epsfxsize=80mm
\epsfysize=67mm
\centerline{\epsfbox{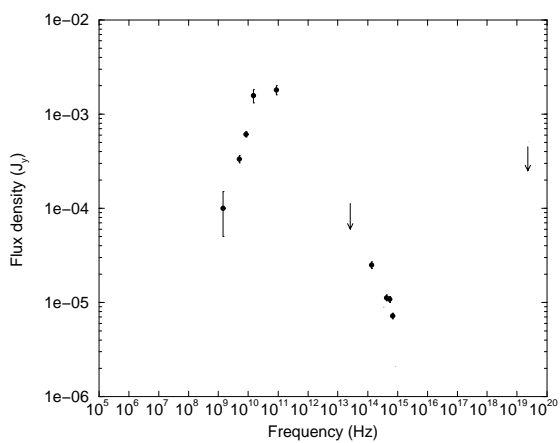}}
\caption{The multiwaveband spectrum of OT/GRB 970508 on $\sim$ 20 May 1997.
Based on observations obtained at the VLA, Ryle, IRAM PdBI, ISO, Keck and 6-m
SAO (Frail et al. 1997, Pooley and Green 1997, Bremer et al. 1998, Hanlon et
al. 1998, Morris et al. 1997 and Sokolov et al. 1998).}
\end{figure}

\section{SUMMARY}

The existence of an X-ray afterglow in {\it all} bursts seems to be confirmed.
The first optical/IR counterparts have been found in 1997. GRB 970508
was also detected in radio and mm, but prompt searches at other 
wavelengths failed to detect GRB 970111, GRB 970402 and GRB 970828. 
 
{\it BeppoSAX} and {\it RossiXTE} have opened a new window in the GRB field and 
it is widely accepted now that most GRBs, if not all, lie at cosmological 
distances. It is expected that {\it BeppoSAX}, {\it RossiXTE} and {\it CGRO} 
will facilitate the discoveries of other counterparts and, together with 
the new high-energy observatories 
({\it AXAF}, {\it SPECTRUM X/$\Gamma$}, {\it XMM}, {\it INTEGRAL}, 
{\it HETE 2}) and other satellites of the future {\it 4th Interplanetary 
Network}, will definitively solve the long-standing Gamma-Ray Burst  
mystery.

\section{ACKNOWLEDGEMENTS}

I am grateful to many colleagues for very fruitful discussions, in particular
to S. Brandt, M. Cervi\~no, E. Costa, M. Feroci, G. Fishman, F. Frontera, 
S. Golenetskii, J. Greiner, J. Gorosabel, S. Guzyi, R. Hudec, K. Hurley, J. 
Isern, C. Kouveliotou, N. Lund, L. Metcalfe, H. Pedersen, M. de Santos, 
A. Shlyapnikov, R. Sunyaev, W. Wenzel, B. Wilson, C. Wolf, W. Wenzel, 
A. Yoshida and many o\-thers. I should also 
mention here the continuous su\-pport of my wife, M. E. Alcoholado-Feltstr\"om, 
who always patiently awaits for my return when sometimes I run off to the 
nearest observatory in order to perform the GRB follow-up observations and 
further 40-hr non-sleeping data analysis.

\end{document}